\documentclass[reprint,superscriptaddress,aip]{revtex4-2}
\usepackage{lineno}\usepackage[normalem]{ulem}
\usepackage{xcolor}\usepackage{tikz}
\usepackage{amsmath,amssymb}
\usepackage{MnSymbol}
\usepackage{graphicx}
\usepackage{dcolumn}
\usepackage{bm}
\usepackage[sort&compress]{natbib}
\usepackage[export]{adjustbox}
\usepackage{xcolor}\usepackage{tikz}
\usepackage{amsmath,amssymb}
\usepackage{MnSymbol}
\usepackage{graphicx}
\usepackage{dcolumn}
\usepackage{bm}
\usepackage[sort&compress]{natbib}
\usepackage[export]{adjustbox}

\usepackage{hyperref}

\newcommand{\rT}{\rho_{\text{T}}}

\newcommand{\vP}{{\bf P}}

 \newcommand{\dx}{\text{d}{x}} \newcommand{\dt}{\text{d}{t}}

\newcommand{\dparf}[2]{\frac{\partial^2 #1}{\partial #2^2}}

\usepackage{pstricks}

\begin{document}

\title{Emergence of rogue-like waves in a reaction-diffusion system: Stochastic output from deterministic dissipative dynamics}

\author{Edgar Knobloch}
\affiliation{Department of Physics, University of California, Berkeley, California 94720, USA}

\author{Arik Yochelis}\email{yochelis@bgu.ac.il}
\affiliation{Swiss Institute for Dryland Environmental and Energy Research, Blaustein Institutes for Desert Research, Ben-Gurion University of the Negev, Sede Boqer Campus, Midreshet Ben-Gurion 8499000, Israel}%
\affiliation{Department of Physics, Ben-Gurion University of the Negev, Be'er Sheva 8410501, Israel}%

\date{\today}

\begin{abstract}
  Rogue waves are an intriguing nonlinear phenomenon arising across different scales, ranging from ocean waves through optics to Bose-Einstein condensates. We describe the emergence of rogue-like wave dynamics in a reaction-diffusion system that arise as a result of a subcritical Turing instability. This state is present in a regime where all time-independent states are unstable, and consists of intermittent excitation of spatially localized spikes, followed by collapse to an unstable state and subsequent regrowth. We characterize the spatiotemporal organization of spikes and show that in sufficiently large domains the dynamics are consistent with a memoryless process.
\end{abstract}

\maketitle
\noindent \textbf{Rogue waves are large excitations that appear intermittently and unpredictably in many conservative systems arising in nature, including nonlinear optics and ocean waves. We show here that in forced dissipative spatially distributed systems, a subcritical Turing bifurcation may also lead to irregular and memoryless rogue-like wave dynamics, i.e., we identify a mechanism whereby deterministic time-dependent dynamics with stochastic properties arise even in the absence of an oscillatory instability. Since the Turing instability is a generic pattern-forming instability of reaction-diffusion models, our results reveal a generic mechanism that sheds fresh light on time-dependent patterns in physicochemical and biological applications.} \\ \\

Rogue waves (RW) are isolated events that ``appear out of nowhere and disappear without a trace'',\cite{AKHMEDIEV2009675} often referred to as freak, monster, or giant waves.\cite{yan2012rogue} RW are usually studied in the context of certain Hamiltonian models~\cite{solli2007optical,Onorato2013,Pelinovsky2016,dudley2019rogue,tlidi2022rogue} describing extreme localized events in oceanic~\cite{dysthe2008oceanic,Pelinovsky2016} and atmospheric flows,\cite{stenflo2010rogue} and systems such as Bose-Einstein condensates,\cite{bludov2009matter} nonlinear optics,\cite{solli2007optical,akhmediev2013recent,dudley2014instabilities,akhmediev2016roadmap,song2020recent,rao2022general} parametrically driven capillary waves,\cite{shats2010capillary} superfluid He,\cite{ganshin2008observation} and even in financial markets.\cite{yan2010financial} Recently, it has been shown that RW may also arise in near-resonant driven dissipative systems,\cite{subramanian2022forced} and in this study, we present a distinct and broadly applicable mechanism for generating RW in a reaction-diffusion system highlighting the role of a subcritical Turing bifurcation.

The Turing bifurcation,\cite{turing1952chemical,dawes2016} and more broadly a finite wavenumber bifurcation,\cite{ch93} is of fundamental importance in the study of steady-state patterns in media driven far from equilibrium, particularly in reaction-diffusion (RD) models.\cite{ch93,maini1997spatial,kondo2002reaction,krause2021introduction} When the bifurcation is supercritical one finds steady small amplitude spatially periodic patterns.\cite{ch93} In the subcritical case, the corresponding small amplitude patterns are unstable but in physically motivated models these states turn around at larger amplitude in a fold (or saddle-node) bifurcation.

\begin{figure}[tp]
    (a){\includegraphics[width=0.45\textwidth]{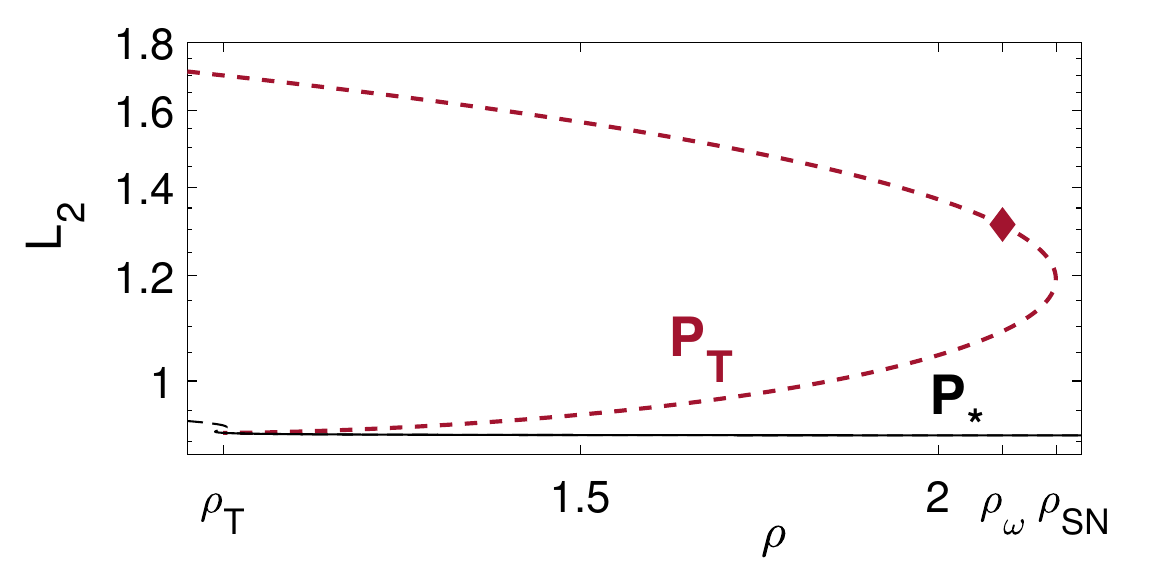}}
    (b)\includegraphics[width=0.45\textwidth]{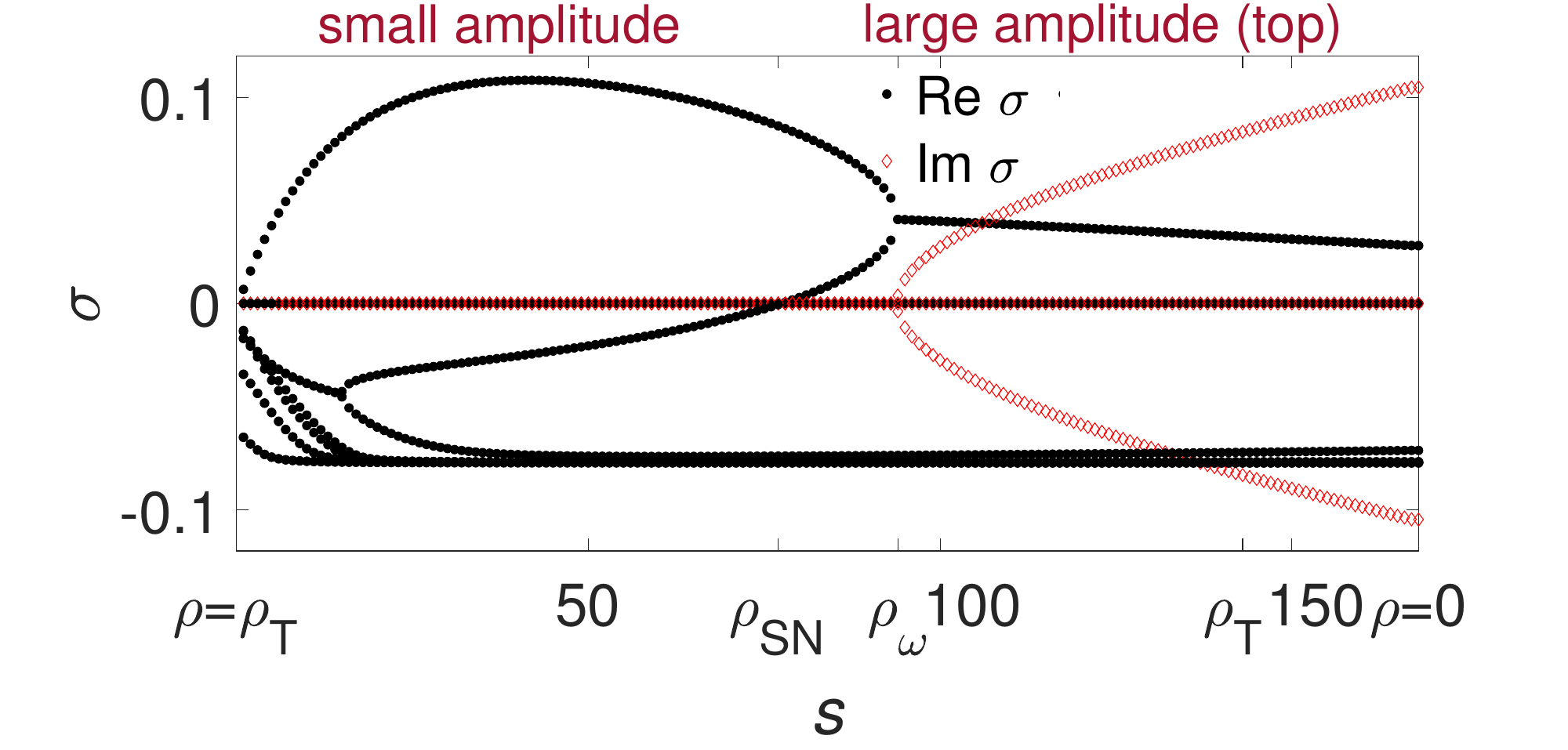}
    \caption{(a) Bifurcation diagram showing the norm ${\text L}_2$ (on a logarithmic scale) of solutions of the model (\ref{eq:AI}) as a function of the parameter $\rho$, focusing on the unstable Turing branch $\vP_{\rm T}$ that bifurcates subcritically (towards the right) from a spatially uniform $\vP_*$ state at $\rho=\rT\approx 1.0011$ and folds back towards the left at $\rho=\rho_{\rm SN}\approx 2.165$. Solid (dashed) lines for $\vP_*$ and $\vP_{\rm T}$ indicate temporal stability (instability) with respect to infinitesimal perturbations. 
    (b) Temporal eigenvalues $\sigma$ as a function of the arclength $s$ along the $\vP_{\rm T}$ branch, computed on a domain of length $L=L_{\rm T}\equiv 2\pi/k_{\rm T}\approx 2.88$, i.e., one Turing wavelength,\cite{knobloch2022instability}, ranging from $\rT$ on the left, past the fold and back to lower values of $\rho$, all the way to $\rho=0$ on the right, showing that $\vP_{\rm T}$ remains unstable throughout. Collision of real unstable eigenvalues leads to complex conjugate eigenvalues at $\rho=\rho_\omega \approx 2.09$ as also indicated by the diamond symbol in (a). Here ${\text L}_2\equiv \sqrt{L^{-1}\int_{0}^{L} \dx [{\sum_{j} P_j^2+(\partial_x P_j)^2}]}$, $P_j=A,H,S,Y$, and $L$ is the domain length.}
\label{fig:uni}
\end{figure}

\begin{figure*}[tp]
	\begin{flushleft}
	(a)\includegraphics[width=0.3\textwidth]{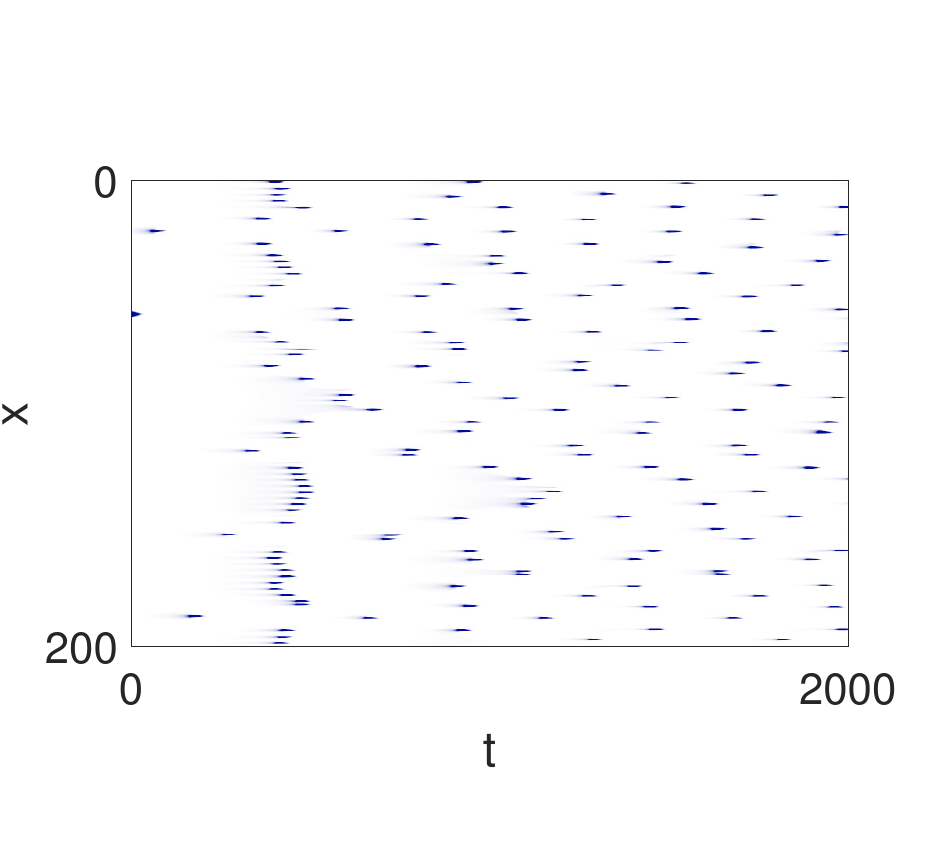}
	\quad(b)\includegraphics[width=0.3\textwidth]{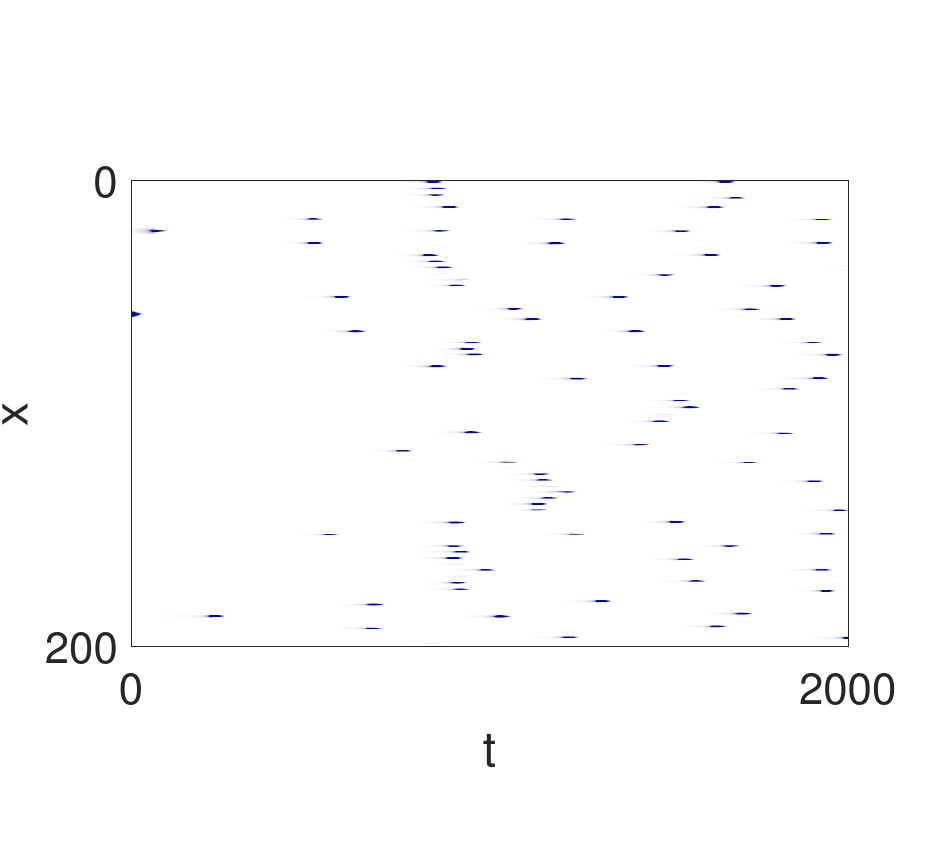}
	(c)\includegraphics[width=0.3\textwidth]{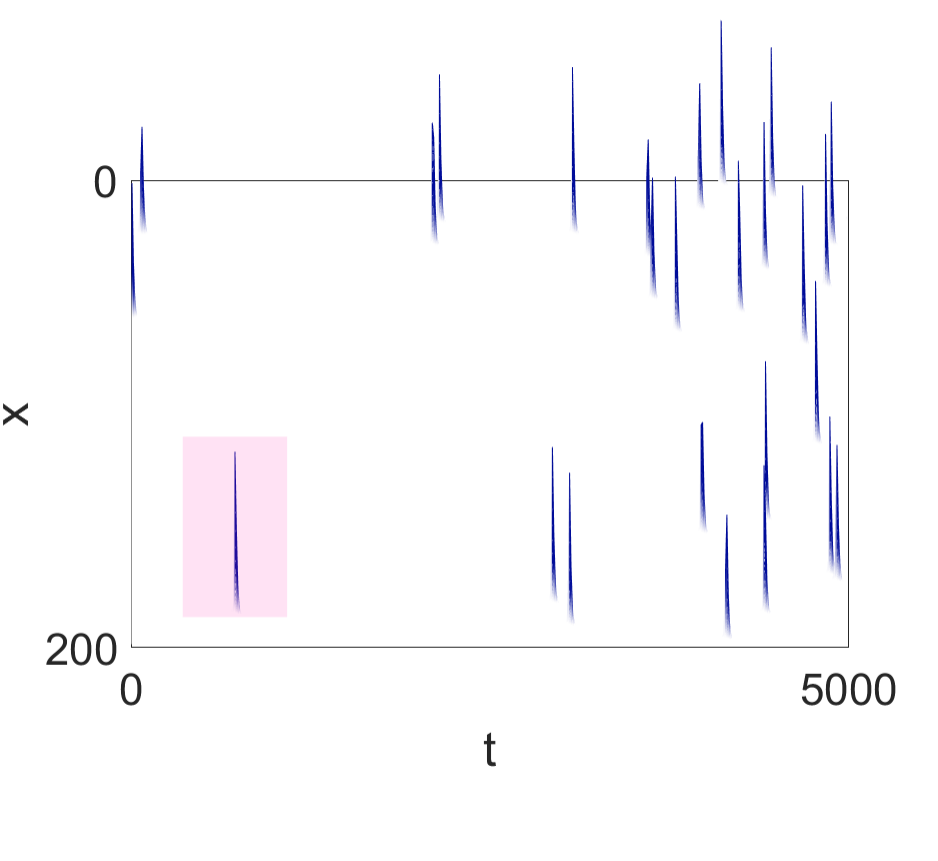}
	\end{flushleft}
        {\includegraphics[height=1.5in]{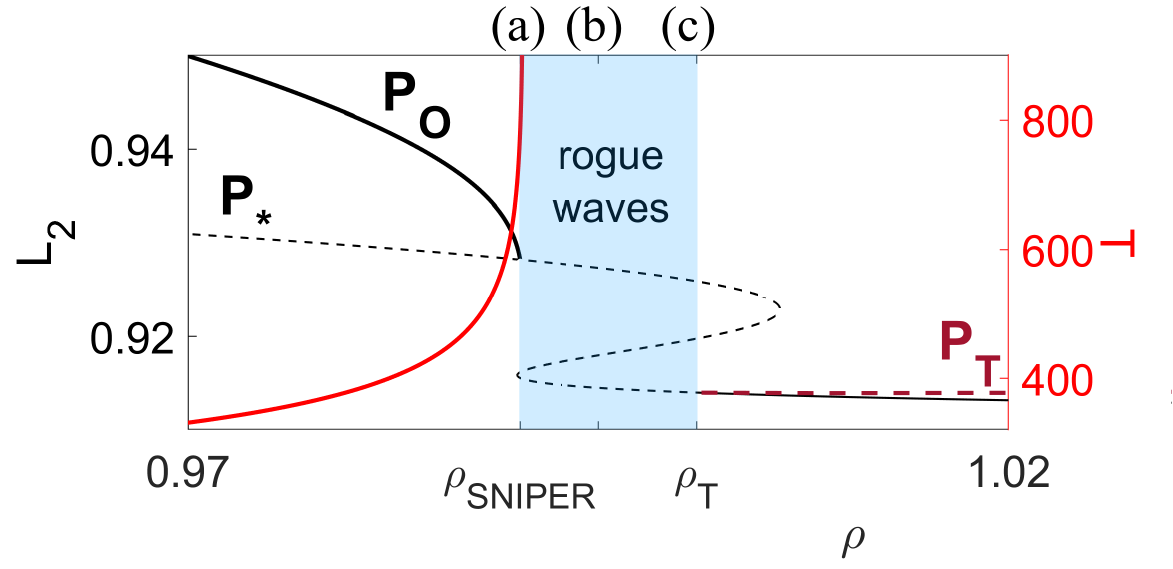}}
        \hskip 0.2in {\includegraphics[height=1.4in]{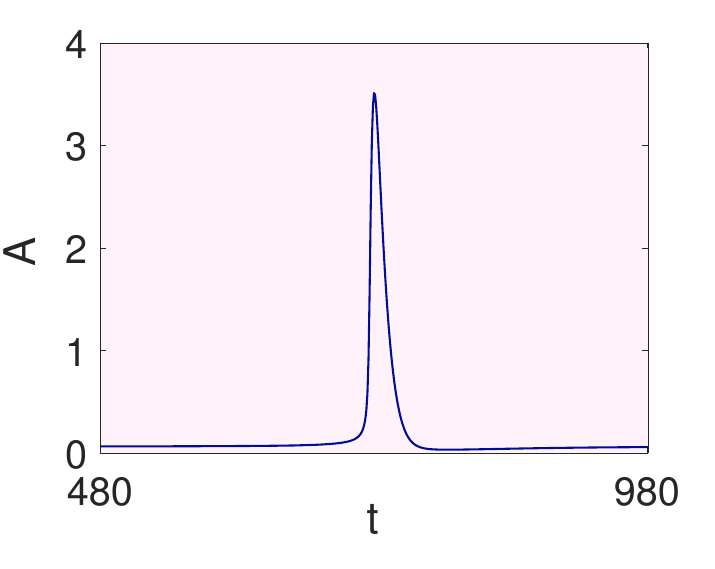}}
        {\includegraphics[height=1.4in]{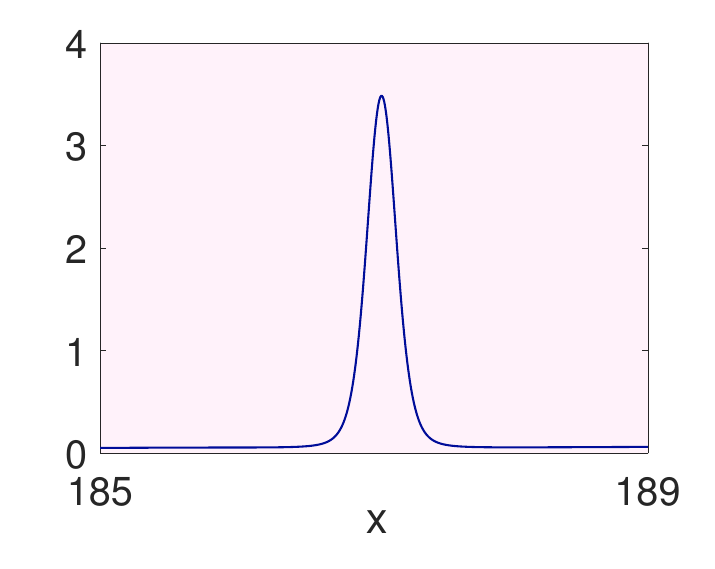}}
        \begin{flushleft}
        (d)\includegraphics[width=0.3\textwidth]{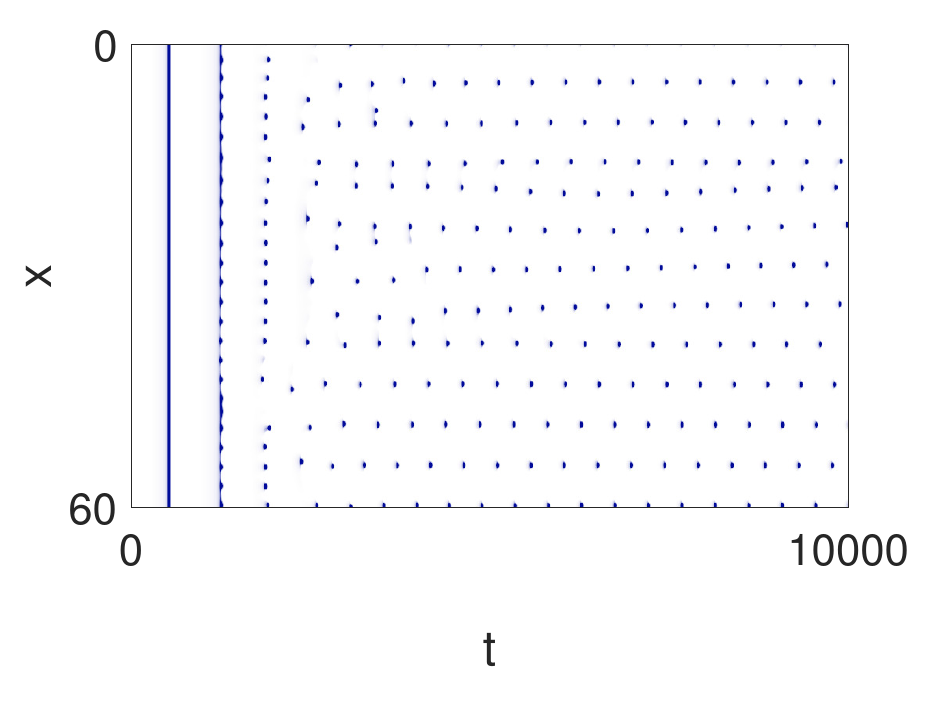}
	\quad(e)\includegraphics[width=0.3\textwidth]{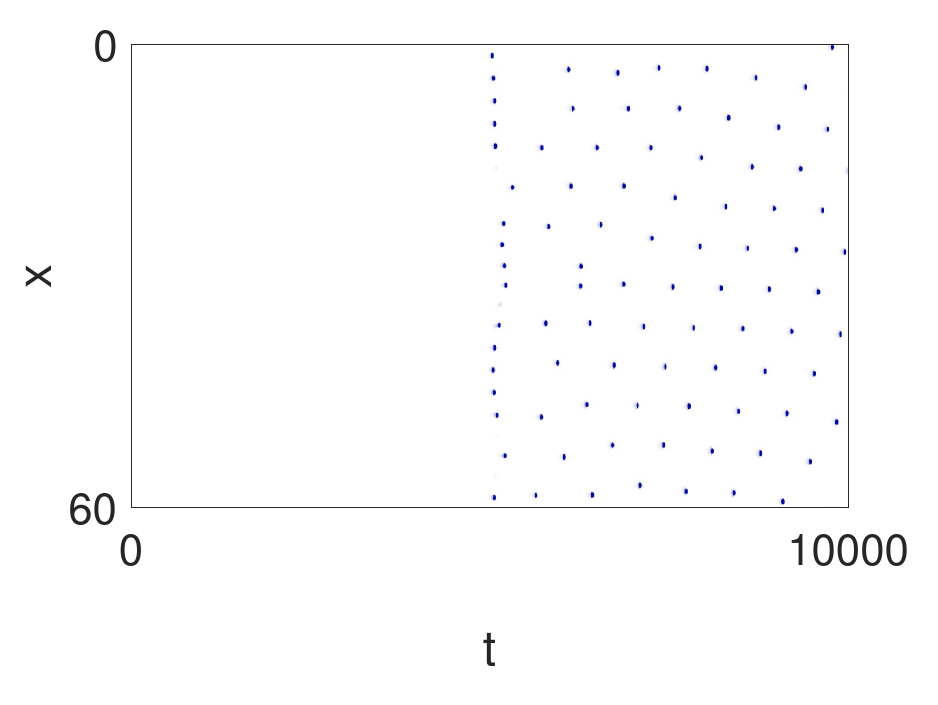}
	(f)\includegraphics[width=0.3\textwidth]{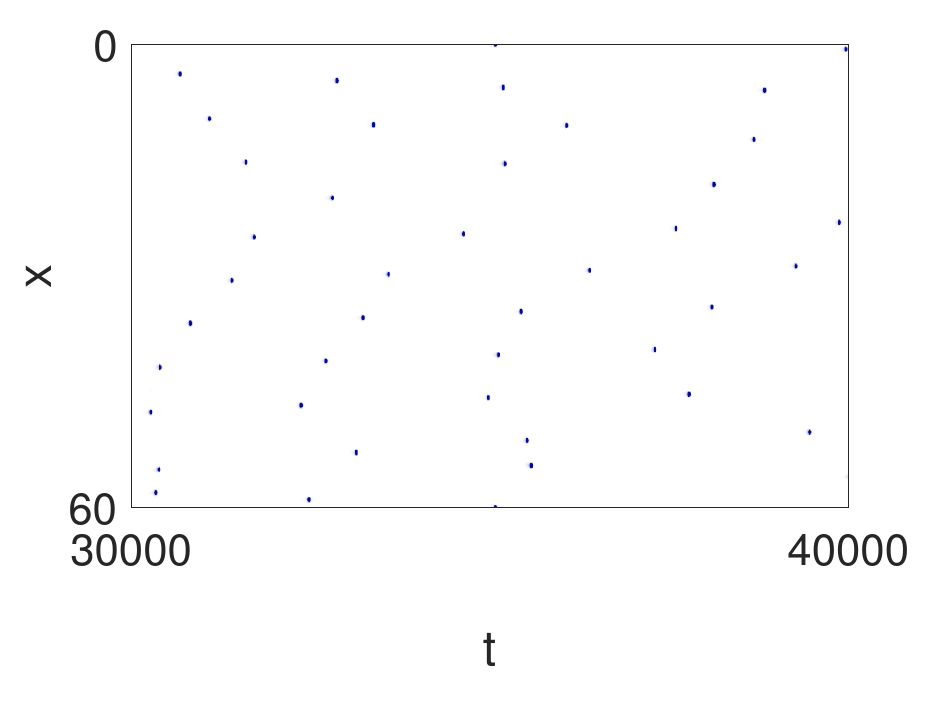}
	\end{flushleft}
        \caption{Summary of the $(\rho,L)$ parameter space together with sample solutions (a-f) shown in space-time plots  of $A(x,t)$ obtained from direct numerical simulations (DNS) of the model (\ref{eq:AI}) with periodic boundary conditions (PBC) in domains of large (a-c) and moderate (d-f) lengths, $L=200$ and $L=60$, respectively. Rogue-like waves (RW) are present in $\rho_{\rm SNIPER}\approx 0.99005\lesssim\rho<\rho_{\rm T}\approx 1.0011$ (left middle panel, blue shaded region). The right middle panels show the temporal and spatial profile of a typical spike taken from panel (c), highlighted in pink. For clarity, in the other panels, we replace the space-time-amplitude spike representation used in (c) with surface plots where darker colors correspond to higher values of $A(x,t)$. For (a-c) the initial condition is a spatially localized perturbation (at about $x=60$, taken from DNS at $\rho=0.998$) embedded in the background corresponding to the lower portion of the $\vP_*$ branch. For (e,f) the initial condition consists of small amplitude random perturbation of $\vP_*$ (lower branch), while for (d) it is the minimum amplitude state from the uniform oscillations present for $\rho<\rho_{\rm SNIPER}$, also close to $\vP_*$. Panels (a,d) are for $\rho=0.99\lesssim \rho_{\rm SNIPER}$, while (b,e) are for $\rho_{\rm SNIPER}<\rho=0.995<\rho_{\rm T}$. Solutions (c,f) are computed at $\rho=1.0\lesssim\rho_{\rm T}$ [in (f) we omit solutions for $t<30000$, in which only negligible fluctuations about $\vP_*$ exist, cf. $t<5000$ in (e)].
        Here ${\text L}_2\equiv \sqrt{(LT)^{-1}\int_0^T \int_{0}^{L} \dt\, \dx [{\sum_{j} P_j^2+(\partial_x P_j)^2}]}$, $P_j=A,H,S,Y$, $L$ is the domain length and $T$ is the temporal period of the $\vP_{\rm O}$ solutions.} 
	\label{fig:x_t_3D}
\end{figure*}

The key question here is: are the large amplitude states beyond the fold stable or do they remain unstable? In one-variable models only the first situation arises. In multi-variable systems, however, this is no longer the case and the pattern state need not gain stability at the fold.\cite{batiste2001simulations} Figure~\ref{fig:uni}(a) shows an example computed from the model discussed below. Here a subcritical Turing branch $\vP_{\rm T}$, created in a bifurcation at $\rT$ as a parameter $\rho$ decreases, remains {\it unstable} even after it folds back. In fact, the large amplitude Turing branch remains unstable even in $\rho<\rT$, i.e. in the parameter regime usually referred to as {\it supercritical}.\cite{yochelis2021nonlinear} The absence of stable Turing patterns (and other steady states) implies that the regime $\rho<\rT$ is necessarily associated with nontrivial time dependence. The instability of the Turing branch is confirmed in Fig.~\ref{fig:uni}(b). The present study focuses on this regime and identifies a mechanism giving rise to intermittent spatially localized spiking that ranges from periodic to {\it rogue-like wave} events. The different regimes of interest and representative spiking oscillations are summarized in Fig.~\ref{fig:x_t_3D}.

\begin{figure}[tp]
	(a)\includegraphics[width=0.45\textwidth]{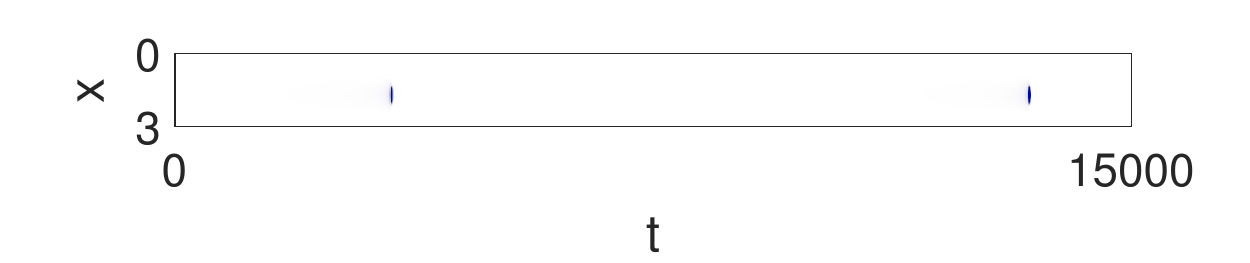}
	(b)\includegraphics[width=0.45\textwidth]{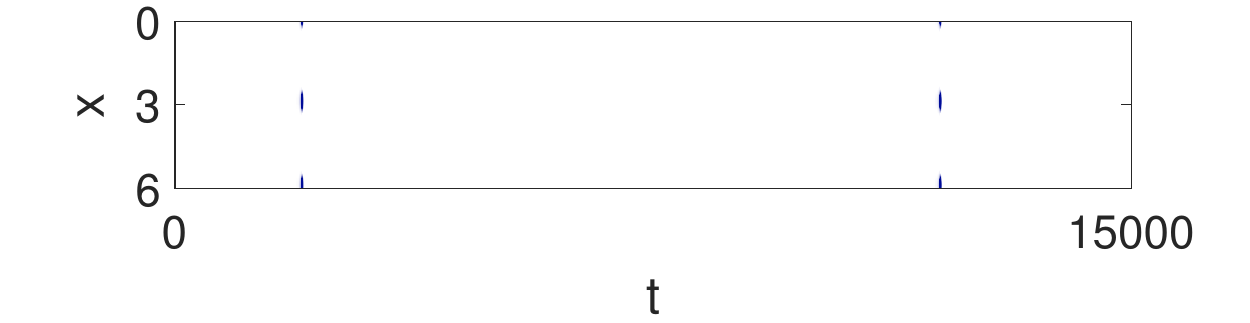}
        (c)\includegraphics[width=0.45\textwidth]{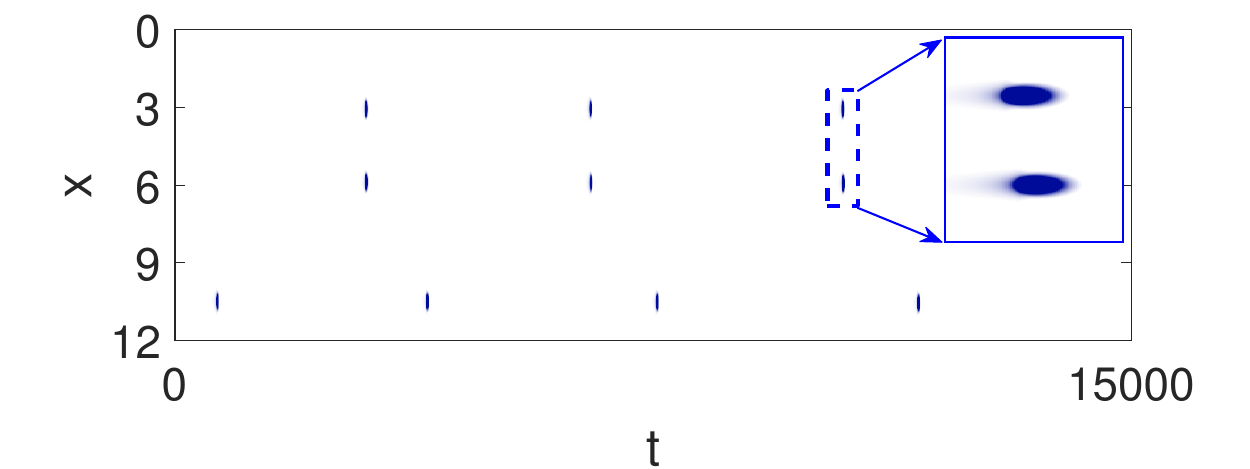}
	(d)\includegraphics[width=0.45\textwidth]{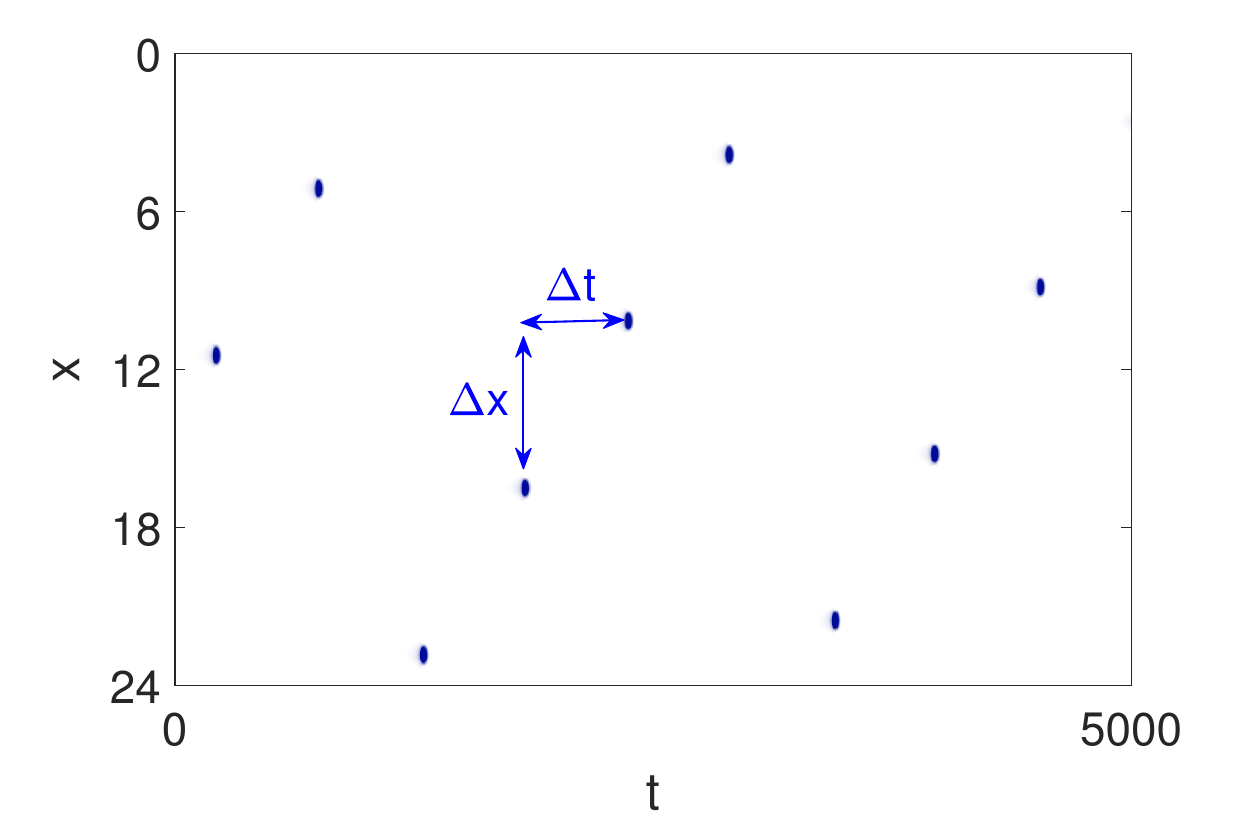}
        (e)\includegraphics[width=0.45\textwidth]{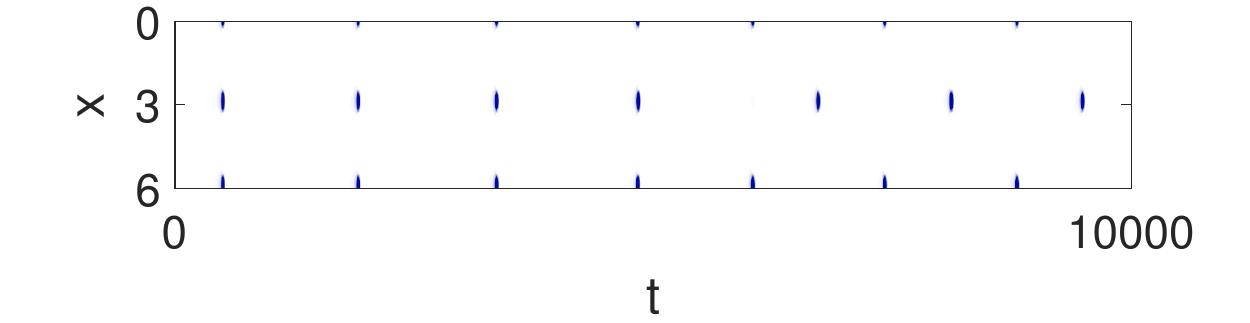}
	\caption{(a-d) Space-time plots of $A(x,t)$ obtained from DNS of \eqref{eq:AI} with PBCs at $\rho=1$, showing periodic dynamics of spikes (a-c) or jumping oscillons (d), depending on the domain length, after decay of transients. The jumps $\Delta t\equiv t_{n+1}-t_{n}$ in time and $\Delta x\equiv |x_{n+1}-x_{n}|$ in space, are indicated in (d). (e) Instability of a synchronous 2-oscillon state at $\rho=0.995$, starting from the profile in (b) at $t=15000$.}
\label{fig:period_L}
\end{figure}
\begin{figure}[ht!]
        (a)\includegraphics[width=0.45\textwidth]{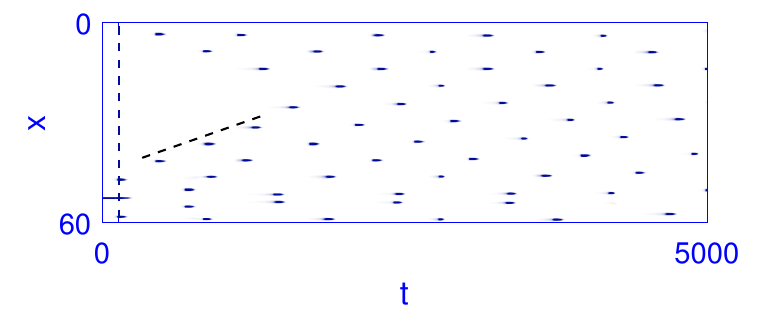}\\
        (b)\includegraphics[width=0.45\textwidth]{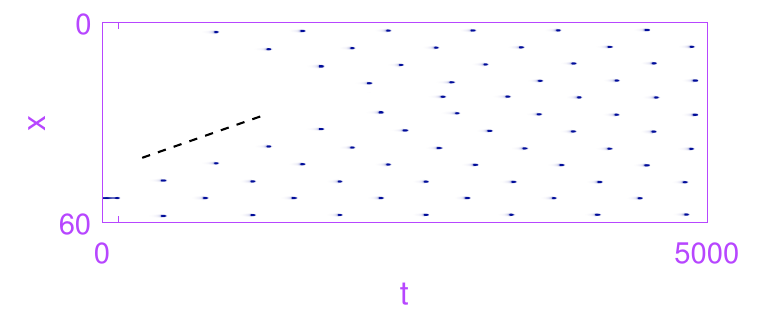}\\
        (c)\includegraphics[width=0.45\textwidth]{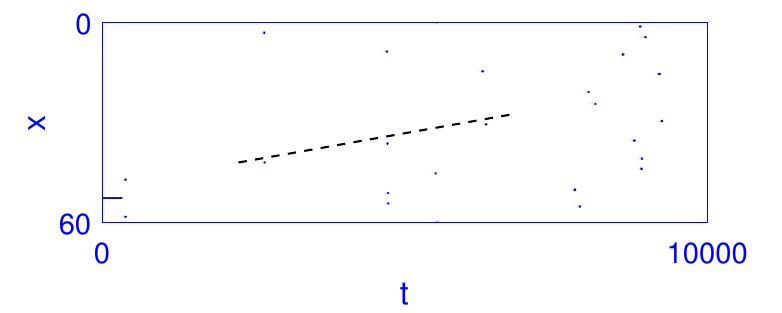}\\
        (d)\includegraphics[width=0.45\textwidth]{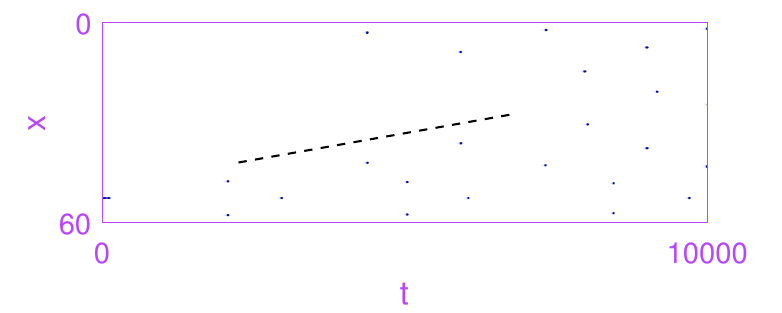}\\
        (e)\includegraphics[width=0.46\textwidth]{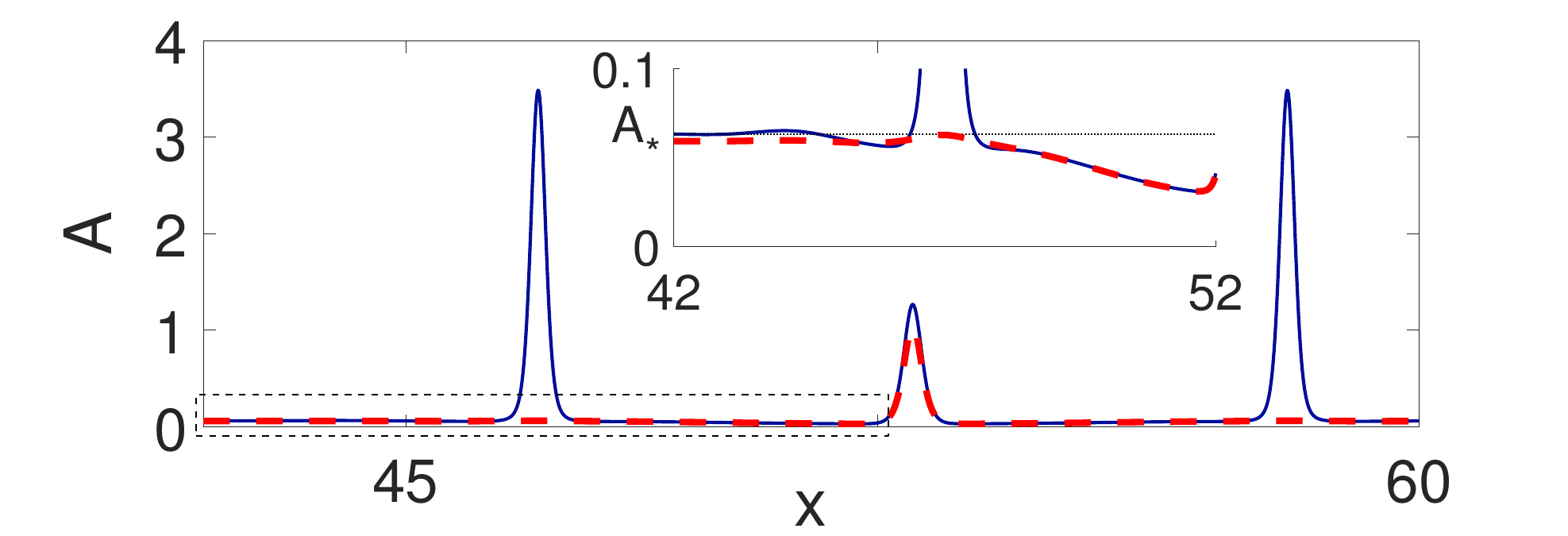}
          \caption{Space-time plots with PBCs at (a,b) $\rho=0.995$ and (c,d) $\rho=1$, initialized with single peak states at (a,c) $\rho=\rT$ (blue panels) and (b,d) $\rho=1.5$ (pink panels), where the former (latter) has nonmonotonic (monotonic) tails; see Ref.\cite{knobloch2021stationary} for details. (e) Spatial profile (in dark blue) at $t=138$ for $\rho=0.995$ along the dashed vertical cut in (a) superposed on the $\rho=\rho_{\rm T}$ initial condition in dashed red; inset shows that the resulting spike profile (blue) is not monotonic.}
\label{fig:evolution_oscillon}
\end{figure}
\begin{figure}[ht!]
        (a)\includegraphics[width=0.47\textwidth]{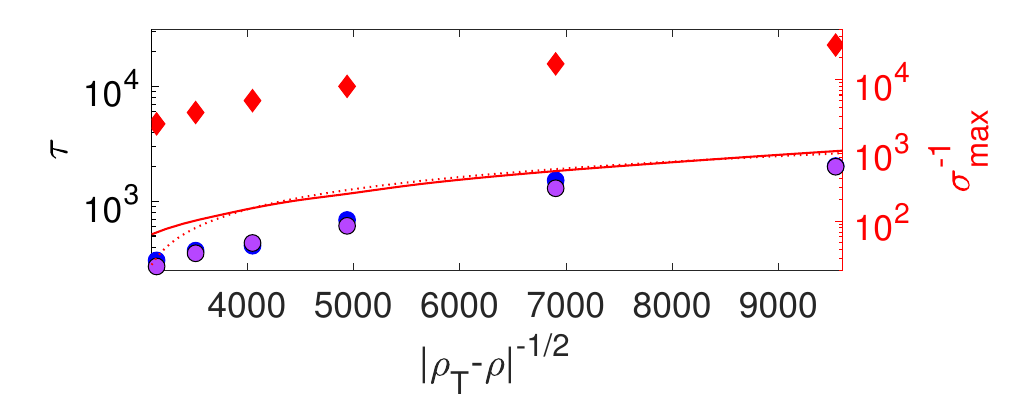}
        (b)\includegraphics[width=0.45\textwidth]{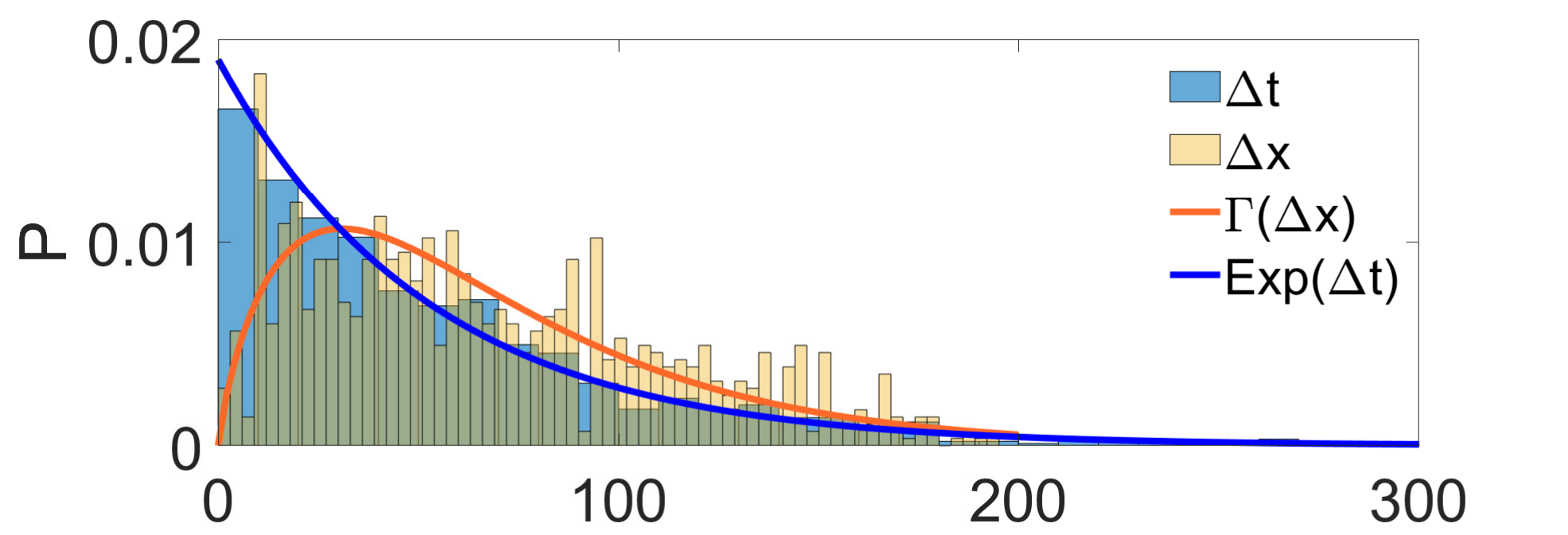}
  \caption{(a) Average jump period $\tau$ of oscillons (color-coded dots, logarithmic scale) along the dashed black lines in Figs.~\ref{fig:evolution_oscillon}(a-d) for $\rho=0.991,0.993,0.995,0.997,0.999,1.0$ showing that $\tau$ is independent of initial conditions and follows the numerically computed inverse of the maximum growth rate $\sigma_{\rm max}(\rho)$ of the Turing instability of $\vP_*$ (solid red curve; the red dotted line is a $|\rT-\rho|^{-1/2}$ fit). The diamonds show the time to the first excitation from a small amplitude perturbation of $\vP_*$, as in Figs.~\ref{fig:x_t_3D}(e,f).
  (b) Probability density function (PDF) of successive spike locations in time ($\Delta t\equiv t_{n+1}-t_{n}$) and in space ($\Delta x\equiv |x_{n+1}-x_{n}|$) when $\rho=1$, computed in an interval of 50000 time units [starting from the last profile in Fig.~\ref{fig:x_t_3D}(c)]. The bin width in time is taken as the width of the spike at about half-height, leading to bins of width 10, while in space is taken as 3, respectively; see Appendix~\ref{appendix} for fit details.}
\label{fig:period_oscillon_prof}
\end{figure}

We employ Meinhardt's four-variable activator-inhibitor-substrate model~\cite{meinhardt1976morphogenesis} in one spatial dimension, where the four fields $A$, $H$, $S$ and $Y$ represent the concentrations of an activator, an inhibitor, the substrate, and a marker for differentiation:\cite{yao2007matrix}
\begin{subequations}\label{eq:AI}
	\begin{eqnarray}
		\frac{\partial A}{\partial t}&=&c\dfrac{SA^2}{H}-\mu A+\rho_{\text{A}} Y+D_{\text{A}} \dparf{A}{x}, \\
		\frac{\partial H}{\partial t}&=&cSA^2-\nu H+\rho_{\text{H}} Y+D_{\text{H}} \dparf{H}{x}, \\
		\frac{\partial S}{\partial t}&=&c_0-\gamma S-\varepsilon Y S+D_{\text{S}} \dparf{S}{x}, \\
		\frac{\partial Y}{\partial t}&=&d A-eY+\dfrac{Y^2}{1+fY^2}+D_{\text{Y}} \dparf{Y}{x}.
	\end{eqnarray}
\end{subequations}
As in earlier work,\cite{yochelis2021nonlinear,knobloch2021stationary,knobloch2022instability,knobloch2023front} we employ the inhibitor control level, $\rho_{\text{H}}$, as a control parameter while keeping all other parameters fixed and within the range of previous studies: $c=0.002$, $\mu=0.16$, $\rho_{\text{A}}=0.005$, $\nu=0.04$, $c_0=0.02$, $\gamma=0.02$, $\varepsilon=0.1$, $d=0.008$, $e=0.1$, $f=10$, $D_{\text{A}}=0.001$, $D_{\text{H}}=0.02$, $D_{\text{S}}=0.01$, with $D_{\text{Y}}=10^{-7}$ to model negligible $Y$ diffusion (in the spirit of the original model). For these parameter values it is convenient to present our results in terms of $\rho\equiv 10^5\rho_H$. 

Figure~\ref{fig:x_t_3D}, middle-left panel, shows the parameter regime of interest, with $\vP_*$ denoting the branch of spatially uniform states, together with the location of the key bifurcation points $\rho=\rho_{\rm SNIPER}\approx 0.99005$ and $\rho=\rT\approx 1.0011$ on this branch. The SNIPER bifurcation~\cite{strogatz2018nonlinear} generates large amplitude, spatially coherent, long period oscillations in $\rho<\rho_{\rm SNIPER}$, labeled $\vP_{\rm O}$, with period $T$ that diverges as $(\rho_{\rm SNIPER}-\rho)^{-1/2}$ near $\rho_{\rm SNIPER}$. Beyond $\rho_{\rm SNIPER}$ no uniform oscillations are present and the system evolves into different spiking states, depending on initial conditions and domain size. We focus on the region $\rho_{\rm SNIPER}<\rho<\rT$ (blue shaded region in Fig.~\ref{fig:x_t_3D}) and explore the development of spiking as a function of the domain size $L$ when $\rho=1$. 

Figure~\ref{fig:period_L}(a) shows periodic spiking of a single {\it oscillon} when $L=3$. Panel (b) shows stable synchronous spiking of two oscillons when $L=6$, while (c) shows that three oscillons in an $L=12$ domain oscillate asynchronously, although their spatial location remains fixed. When $L=24$ the spikes reorganize into a {\it jumping oscillon} (JO) state,\cite{yang2006jumping,cherkashin2008discontinuously,knobloch2021origin} i.e., a periodically oscillating and translating state, as demonstrated in (d). Panel (e) shows that the synchronous 2-oscillon state is unstable at $\rho=0.995$. The JO in Fig.~\ref{fig:period_L}(d) travels in the direction of decreasing $x$, with jump distance $\Delta x\equiv |x_{n+1}-x_n| \sim 6$ and time $\Delta t \equiv t_{n+1}-t_n \sim 550$ between jumps, and reemerges at $x=0$ after a transit time of order $2500$, corresponding to translation speed $c\sim 0.01$. Here $x_n$, $t_n$ denote the location and time of the $n^{\rm th}$ spike as determined by its maximum.

Such JOs can evolve from small perturbations of the unstable $\vP_*$ state or from finite amplitude spatially localized initial conditions as in Figs.~\ref{fig:evolution_oscillon}(a-d). These space-time plots show the initial behavior for different $\rho<\rT$ when an $L=60$ system is initialized with a steady but unstable peak from $\rho>\rT$.\cite{knobloch2021stationary} In each case this state necessarily collapses but in doing so it emits jumping oscillons that propagate outwards with a well-defined speed that depends sensitively on $\rho$. In (e) we show that the intermittent spikes achieve a larger amplitude than the steady peaks present in $\rho>\rT$ used to initialize the computation and that these are triggered by small inhomogeneities in the initial condition (dashed curve). The spatial profile of the spikes depends on the parameter $\rho$ since this quantity determines the spatial eigenvalues $\lambda$ of $\vP_*$ that are reflected in the tail of the spikes.\cite{knobloch2021origin} 

In general, the JOs organize in an approximately standing pattern for smaller values of $\rho$ and $L$, with larger domains necessary to eliminate the influence of the domain size and the influence of the initial conditions. When $L=60$ the JOs tend to form wavefronts consisting typically of six oscillons in the domain, although the oscillons do not oscillate in synchrony. In contrast, for $\rho$ sufficiently close to $\rho_{\rm T}$ and/or larger $L$ the resulting intermittent spiking consists of short duration, well-localized spikes at apparently random locations and times [see Figs.~\ref{fig:x_t_3D}(a-c)]. The mean interspike interval is an increasing function of $\rho$ as expected from the positive growth rate $\sigma_{\rm max}(\rho)\equiv{\rm max}_k \sigma(k,\rho)$ of the Turing instability of $\vP_*$, maximized over perturbation wave numbers $k$, and by $\rho=1$ the spiking occurs only rarely and the spatial location of the spikes is entirely unpredictable. This state exhibits all the hallmarks of a {\it rogue wave} state, except for the fact that the spikes emerge from (and collapse back into) a linearly unstable state instead of a small amplitude wave-like state, and are all of essentially uniform height, determined by a balance between forcing, nonlinearity and dissipation, i.e., by the value of $\rho$. Both effects are a consequence of the forced-dissipative nature of the system~\eqref{eq:AI}. 

Figure~\ref{fig:period_oscillon_prof}(a) shows the jump period $\tau$ of the emitted oscillons until such time that the counterpropagating JOs begin to interact, and compares it with the inverse of the maximum growth rate $\sigma_{\rm max}(\rho)$ of the Turing instability of $\vP_*$ (red curve); since the corresponding wave number $k_{\rm max}$ approaches the Turing wave number $k_{\rm T}$ linearly as $\rho\to\rT$, this time scale is expected to follow $\sigma^{-1}_{\rm max}\sim|\rho-\rT|^{-1/2}$, as is indeed the case. Thus $\tau$ grows rapidly with $\rho$ reflecting the decreasing instability growth rate as $\rho$ approaches $\rT$ from below. This is in contrast to the usual behavior observed in excitable systems where the JO propagation speed does not exhibit sensitive dependence on parameters. The resulting front-like propagation of the spiking behavior resembles that arising in the pearling instability~\cite{goldstein} or the Eckhaus instability,\cite{ma_knobloch2012} propagative behavior that is also generated by a real eigenvalue.

To characterize the RW regime, we computed the one-dimensional probability density functions (PDFs) $P(\Delta x)$ and $P(\Delta t)$. These PDFs are generated by binning the data as described in Fig.~\ref{fig:period_oscillon_prof}(c) and computed at $\rho=1.0\lesssim \rho_{\rm T}$ with $L=200$. The latter PDF is well fitted (regression coefficient $R^2=0.9929$) by an exponential function (see Appendix~\ref{appendix} for details), indicating that RW generation is a Poisson or memoryless random process. On the other hand, $P(\Delta x)$ is well approximated (regression coefficient $R^2=0.9458$) by a $\Gamma$ distribution (see Appendix~\ref{appendix}) as observed for RW over varying topography.\cite{bolles2019anomalous} This distribution works well for $\Delta x\gtrsim30$ but greatly underestimates separations near the favored separation $\Delta x\approx 10$ [Figs.~\ref{fig:evolution_oscillon}(a,b)]. We conclude that in large domains spiking occurs at a fixed parameter-dependent mean rate in time subject to lateral inhibition in space. 

In this study, we have demonstrated that in dissipative, spatially distributed systems a (subcritical) Turing bifurcation may lead to dynamics resembling rogue waves, i.e., large amplitude waves that appear intermittently and unpredictably in time and space. These states arise in the Turing-unstable (supercritical) regime (here $\rho<\rho_{\rm T}$) whenever the Turing branch ($\vP_{\rm T}$) remains unstable in this regime~(Fig.~\ref{fig:x_t_3D}). In sufficiently large domains these irregular but deterministic events resemble a memoryless stochastic process.\cite{yeung1998nonlinear,erban2009analysis} Given the exponential sensitivity of the spiking process to the details of the small amplitude interpulse behavior, we anticipate that in the presence of small amplitude spatio-temporal noise, systems of this type will exhibit all the hallmarks of a rogue-wave state. Moreover, since the Turing instability is a generic pattern-forming instability of activator-inhibitor models,\cite{ch93,maini1997spatial,kondo2002reaction,krause2021introduction} our results identify a new generic mechanism leading to persistent temporal dynamics that does not rely on the presence of an oscillatory instability or noise.\cite{wacker1995transient,tel2008chaotic,tlidi2022rogue} The study thus broadens the class of systems in which stochastic dynamics may arise out of deterministic models, whether of fluid tubulence or active nematics.\cite{gil1990statistical,avila2023transition,doostmohammadi2017onset,reinken2022ising}

We expect this mechanism to operate in quasi-one-dimensional systems obeying biochemical kinetics, ranging from the initiation of developmental signaling~\cite{affolter2009tissue,iber2013control,davies2015epithelial} to intracellular waves,\cite{Allard2013,beta2023actin} while broadening the range of possible mechanisms governing complex spatiotemporal oscillations in biophysical systems such as spontaneous otoacoustic emission in the cochlea~\cite{shera2022whistling} and interspike intervals in integrate-and-fire neural circuits.\cite{levenstein_okun2023} Indeed, similar dynamics may have been observed in other spatially extended systems,\cite{beaume2018three,laing2023,shashangan2024mitigation} albeit in much smaller domains.

\begin{acknowledgments}
\noindent We are grateful to A. van Kan and N. Smith for helpful comments. This work was supported in parts by the National Science Foundation under grant DMS-1908891 (EK), by the Israel Science Foundation under grant 1224/21 (AY), and by grant 2022072 from the United States - Israel Binational Science Foundation (BSF), Jerusalem, Israel (EK and AY).
\end{acknowledgments}

\section*{AUTHOR DECLARATIONS}
\noindent \textbf{Conflicts of Interest.} The authors have no conflicts to disclose.

\section*{DATA AVAILABILITY}
\noindent The data that support the findings of this study are available
from the corresponding author upon reasonable request.
\\
\appendix
\section{Fitting of distributions}\label{appendix}
\begin{figure}[tp]
\centering
(a)\includegraphics[width=\columnwidth]{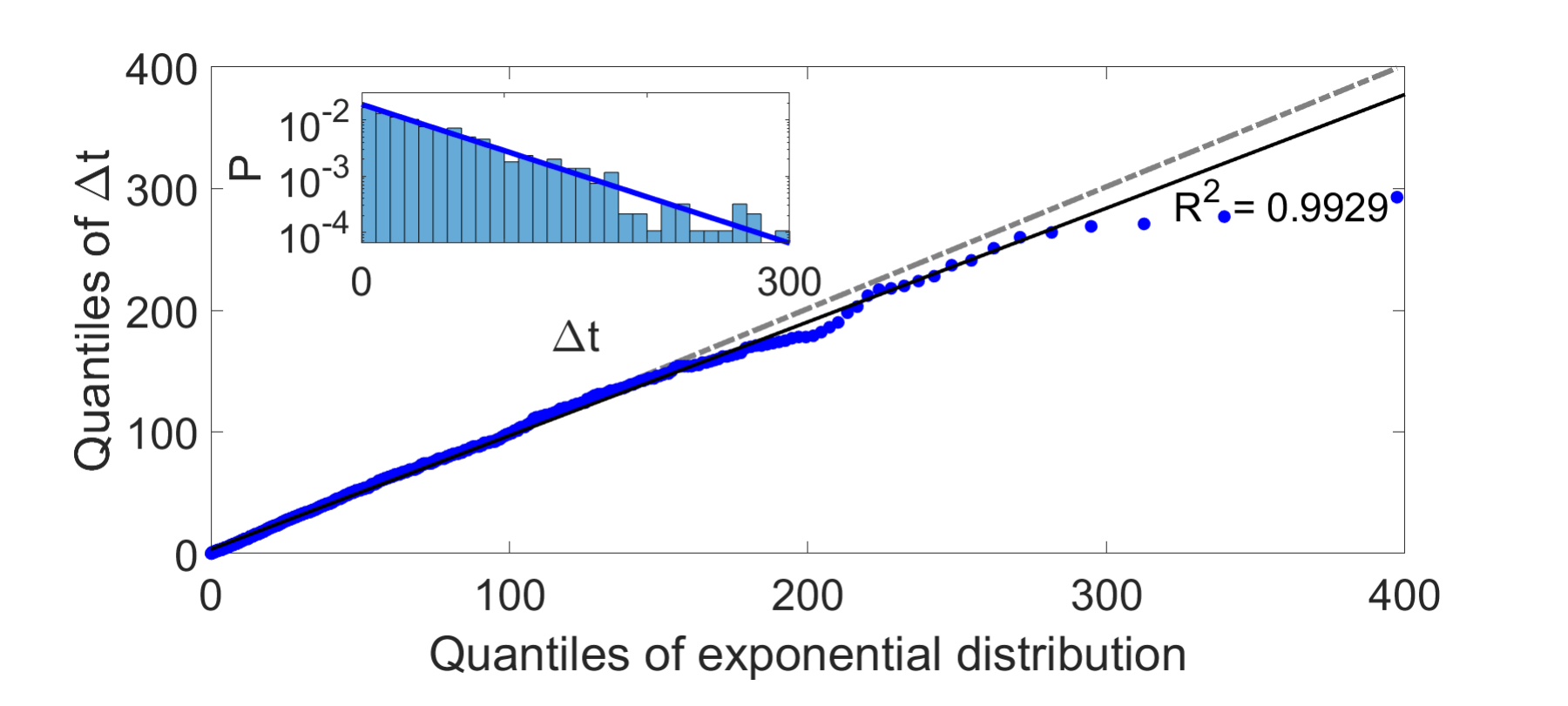}
(b)\includegraphics[width=\columnwidth]{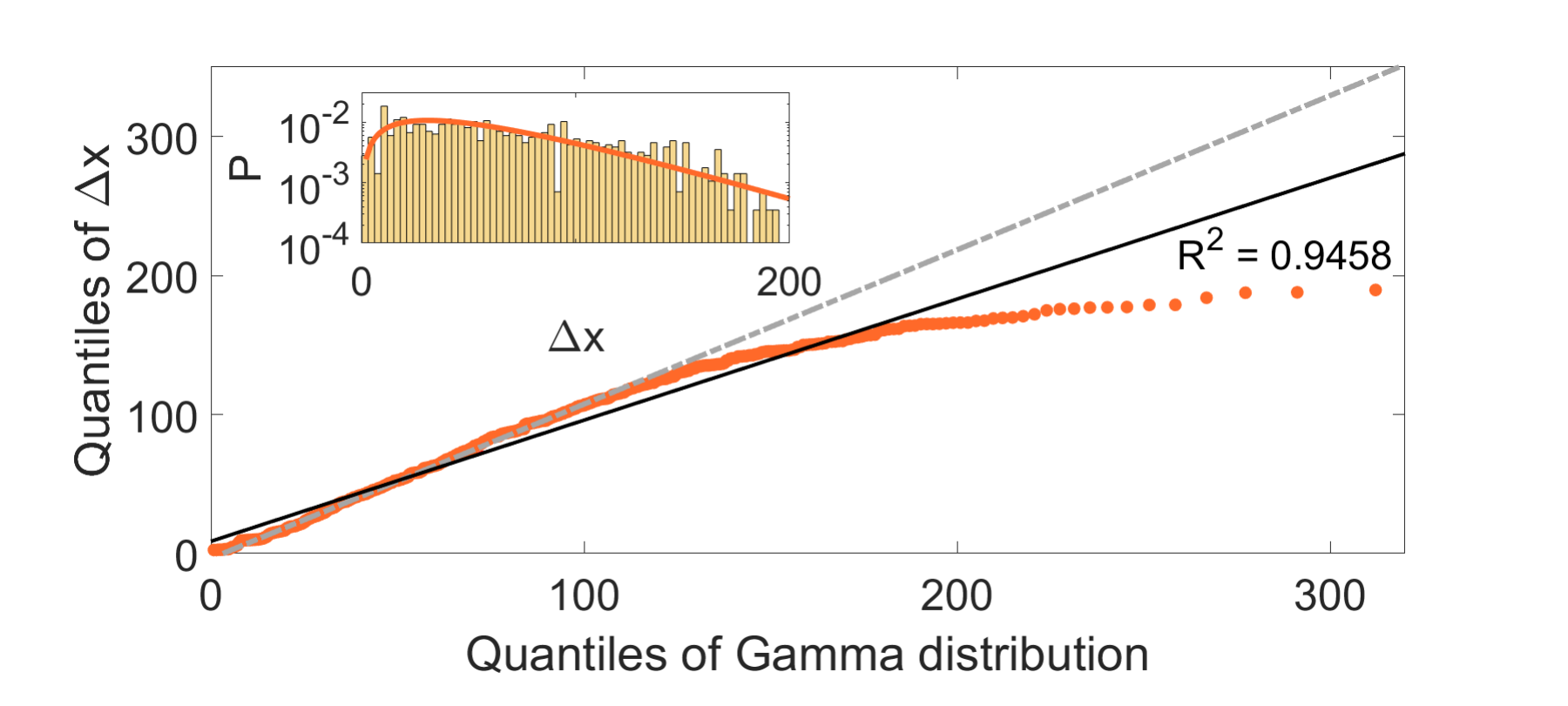}
\caption{Quantiles of (a) $\Delta t$ and (b) $\Delta x$ for the exponential and Gamma ($\Gamma$) distributions (in $\bullet$), respectively, while the insets show the PDFs [from Fig. 5(b)] on a logarithmic scale. Dashed lines represent perfect distributions while solid lines represent a linear fit with regression coefficient $R^2$.}
\label{fig:fit}
\end{figure}
\noindent For fitting of the probability density function (PDF) we use two forms, as shown in Fig.~\ref{fig:fit}:\\
The exponential distribution
\begin{equation}\label{eq:exp_dist}
    \rm{P}(\Delta t;\eta)=\frac{1}{\eta}e^{-\Delta t/\eta},
\end{equation}
and the Gamma distribution
\begin{equation}\label{eq:gamma_dist}
    \rm{P}(\Delta x;\alpha,\beta)=\frac{1}{\Gamma(\alpha)\beta^\alpha}\Delta x^{\alpha-1}e^{-\Delta x/\beta},
\end{equation}
where $\Gamma(\alpha)\equiv \int_0^\infty z^{\alpha-1}e^{-z}\text{d}z$ and $\eta=52.6484$, $\alpha=1.82757$, $\beta=37.3291$.

\section*{REFERENCES}
%

\end{document}